\documentclass[apjpt4]{aastex}
\usepackage{graphicx}
\pdfoutput=1
\begin{document}

\title{Extremely High Excitation SiO Lines in Disk-Outflow System in Orion Source~I}

\author{Mi Kyoung Kim\altaffilmark{1,2}, Tomoya Hirota\altaffilmark{3,4}, Masahiro N. Machida\altaffilmark{5}, Yuko Matsushita\altaffilmark{5},  Kazuhito Motogi\altaffilmark{6}, Naoko Matsumoto\altaffilmark{3,7} \& Mareki Honma\altaffilmark{1,8}}
\email{mikyoung.kim@nao.ac.jp}
\altaffiltext{1}{Mizusawa VLBI Observatory, National Astronomical Observatory of Japan, Hoshigaoka 2-12, Mizusawa-ku, Oshu, Iwate 023-0861, Japan}
\altaffiltext{2}{Korea Astronomy and Space Science Institute, Hwaam-dong 61-1, Yuseong-gu, Daejeon, 305-348, Republic of Korea}
\altaffiltext{3}{Mizusawa VLBI Observatory, National Astronomical Observatory of Japan, Osawa 2-21-1, Mitaka, Tokyo 181-8588, Japan}
\altaffiltext{4}{Department of Astronomical Sciences, SOKENDAI (The Graduate University for Advanced Studies), Osawa 2-21-1, Mitaka, Tokyo 181-8588, Japan}
\altaffiltext{5}{Department of Earth and Planetary Sciences, Faculty of Sciences, Kyushu University, Motooka 744, Nishi-ku, Fukuoka, Fukuoka 819-0395, Japan}
\altaffiltext{6}{Graduate School of Science and Engineering, Yamaguchi University, Yoshida 1677-1, Yamaguchi, Yamaguchi 753-8511, Japan}
\altaffiltext{7}{The Research Institute for Time Studies, Yamaguchi University, Yoshida 1677-1, Yamaguchi, Yamaguchi 753-8511, Japan}
\altaffiltext{8}{Department of Astronomical Sciences, SOKENDAI (The Graduate University for Advanced Studies), Hoshigaoka2-12, Mizusawa-ku, Oshu, Iwate 023-0861, Japan}

\begin{abstract}
{We present high-resolution images of the submillimeter SiO line emissions of a massive young stellar object Orion Source I using the Atacama Large Millimeter/Submillimeter Array (ALMA) at band 8. 
We detected the 464~GHz SiO $v$=4 $J$=11-10 line in Source I, which is the first detection of the SiO $v$=4 line in star-forming regions, together with the 465~GHz $^{29}$SiO $v$=2 $J$=11-10 and the 428~GHz SiO $v$=2 $J$=10-9 lines with a resolution of 50 AU. The $^{29}$SiO $v$=2 $J$=11-10 and SiO $v$=4 $J$=11-10 lines have compact structures with the diameter of $<$80 AU. The spatial and velocity distribution suggest that the line emissions are associated with the base of the outflow and the surface of the edge-on disk. In contrast, SiO $v$=2 $J$=10-9 emission shows a bipolar structure in the direction of northeast-southwest low-velocity outflow with $\sim$200~AU scale. The emission line exhibits a velocity gradient along the direction of the disk elongation. With the assumption of the ring structure with Keplerian rotation, we estimated the lower limit of the central mass to be 7$~M_{\odot}$ and the radius of 12~AU$< r <$26~AU.  
}
\end{abstract}


\section{Introduction}
Orion Source I is a high-mass protostar candidate located in the Orion~KL~(Kleinmann-Low) region \citep{Kleinmann1967}. Source I is thought to have a high-luminosity of $>$10$^4 L_{\odot}$, however, it is so highly embedded that cannot be observed directly at optical and infrared wavelengths.
Due to its proximity~(D$\sim$420~pc)~\citep{Menten2007, Hirota2007, Kim2008}, Source I has been studied in detail via the high-resolution observations with radio and sub-mm wavelengths. 

The source is associated with a low-velocity~($V\sim$18~km~s$^{-1}$) bipolar outflow with a size of $\sim$1000~AU along the northeast-southwest direction \citep{Plambeck2009, Zapata2012,Greenhill2013}. 
The radio continuum observations with Very Large Array (VLA) and ALMA have imaged the elongated continuum emission at the center of the outflow, which is consistent with an edge-on disk perpendicular to the bipolar outflow \citep{Reid2007, Goddi2011,Hirota2015, Plambeck2016, Orozco2017}.
The bright SiO maser emission, which is rare in the star-forming regions, has been found toward Source I.
The high-resolution observations with VLBI revealed that the 43~GHz $v$=1 and 2 $J$=1-0 SiO maser spots are distributed in X-shaped region with a size of $\sim$100~AU  \citep{Greenhill1998, Kim2008, Matthews2010,Issaoun2017}. 
The observations revealed that the masers have a stable velocity gradient in the direction parallel the elongation of the radio continuum, while moving outward along the arms of X-shape. These behaviors suggest that the masers are associated with the rotating outflow from the surface of the disk.
Recent ALMA observations detected high-frequency lines of H$_2$O, CO, and silicon- and sulfur-rich molecules at the position of Source I \citep[e.g.][]{Hirota2012,Hirota2014, Hirota2016a, Hirota2017,Plambeck2016, Ginsburg2018}, which are tracing the disk and the outflowing gas from the disk. The velocity structure of the high-frequency lines also shows the rotational feature similar to those of the 43~GHz $v$=1 and 2 SiO masers. The observed velocity gradients can be modeled with Keplerian rotation curve with the enclosed mass of $\sim$5-9$M_{\odot}$ \citep{Matthews2010, Kim2008, Hirota2014, Plambeck2016, Hirota2017}. 
More recently, \citet{Ginsburg2018} reported the continuum and spectral line emission from Source I's disk and outflow with ALMA observation at a resolution of 0.03-0.06\arcsec. They detected the continuum and several unidentified line emission in the disk, as well as thermal H$_2$O 5$_{5,0}$-6$_{4,3}$ line tracing upper and lower boundary of the disk. The kinematics and morphology of the continuum and line emission confirmed the idea of rotating outflow from the disk suggested in the previous studies, although they obtained the higher central mass of $\sim$15$M_{\odot}$. 

The high-resolution observations of the masers and the high-frequency line emissions reveal a rotating, wide-angle inner outflow and a perpendicular disk around Source I, which can be explained by magneto-centrifugal disk wind launched by the central high-mass young stellar object \citep[e.g.][]{Matthews2010,Greenhill2013,Hirota2017}. This picture is similar to the the low-mass star formation scenario, which describes the transfer of mass and angular momentum between the accretion disk and outflows~\citep{Hirota2017,Vaidya2013}. Besides, the ALMA observations with CH$_3$CN and CH$_3$OH lines revealed the signature of the infall motion toward Source I, which also supports the star formation via accretion \citep{Wu2014}. On the other hand, the proper motion measurements of young stellar objects in Orion KL suggested that dynamical interactions between massive stars played a role in the evolution of Source I \citep{Rodriguez2005,Bally2011,Goddi2011, Rodriguez2017}. 
The scenarios suggest that the dynamical decay of a multiple-star system consist of BN, Source I and perhaps source n occurred $\sim$500 years ago, and released gravitational energy during the interaction would have triggered the explosive outflow seen in H$_2$ and CO. The scenarios expect the Source I to be $\sim 20~M_{\odot}$ binary system \citep{Goddi2011, Bally2011, Rodriguez2017}. The mass of Source I reported by \citet{Ginsburg2018} of 15$M_{\odot}$ is more consistent with these scenarios than those of previous values, although there are still controversies in the detailed process \citep{Chatterjee2012, Luhman2017}.

The study of the physical properties of Source I requires the information of inner outflow/disk region with higher resolution observations. Toward this goal, the high-excitation transition lines could be a potential tool to investigate the hot and dense gas near the protostar \citep{Hirota2014, Hirota2017, Plambeck2016, Ginsburg2018}.
In this paper, we present the results of ALMA observations at 0.1\arcsec~resolution to study dynamical properties of the disk-outflow system in the close vicinity to Source I. The observation yields the maps of SiO lines with the excitation temperature of 3600-7000~K, which trace the inner part of the disk-outflow system of Source I. 

\section{Observations}
Observations were executed on August 27 (460~GHz), and September 22 (430~GHz) in 2015 with the Atacama Large Millimeter/Submillimeter Array (ALMA) as one of the science projects in cycle~2 (ADS/JAO.ALMA\#2013.1.00048.S). Details of observations are summarized in \citet{Hirota2016b, Hirota2017}. The tracking center position of the target source was RA(J2000)=05h35m14.512s, Decl(J2000)=$-$05$^{\circ}$22\arcmin30.57\arcsec.  
The total on-source integration times of the target source were 458~seconds, and 2065~seconds for 460~GHz, and 430~GHz, respectively. 
The line information presented in this paper are summarized in Table \ref{tab-line}. 
For the observations at both 430 and 460~GHz, a primary flux calibrator was J0423$-$013, and a band-pass calibrator was J0522$-$3627. The secondary gain calibrators were J0607$-$0834 for 460 GHz observations and J0501-0159 for 430~GHz observations.
The array was composed of 40 and 35 antennas with the maximum baseline length of 1575~m and 2270~m at 460~GHz and 430~GHz, respectively.
The average system noise temperature was $\sim$300~K at 460~GHz and 500~K at 430~GHz. 
The spectral window centered at 428.0~GHz and 464.6~GHz included observed SiO line emissions. The total bandwidth of the spectral bands was 1875~MHz, and the channel spacing was 976.562~kHz, corresponding the velocity resolution of $\sim$0.6~km~s$^{-1}$. 
 
The self-calibration and the synthesized imaging were carried out using the CASA software package.
The details of the data analysis are summarized in \citet{Hirota2016b, Hirota2017}.
The synthesized images of each velocity channels were made for all lines by using the CASA task {\tt{clean}} with uniform-weighting. The synthesized beam sizes were 0.13\arcsec $\times$0.11\arcsec with the position angle of 82\degr for the 465~GHz $^{29}$SiO $v$=2 $J$=11-10 line, 0.12\arcsec $\times$0.11\arcsec with the position angle of 87\degr for the 464~GHz SiO $v$=4 $J$=11-10 line, and 0.09\arcsec $\times$0.07\arcsec with the position angle of 82\degr for the 428~GHz SiO $v$=2 $J$=10-9 line. 
We made the moment maps using the CASA task {\tt{immoments}}. To make a position centroid maps of the line emission, we did a two-dimensional Gaussian fitting to each velocity channel maps using the task JMFIT with the AIPS software package.
As discussed in \citet{Hirota2016b}, the continuum peak positions at different frequencies deviate from each other by 100~mas at maximum for astrometric calibration errors in the observations or data analysis. Thus, we did not use the absolute position of the emissions in comparing the maps at different frequency bands in this paper.

\begin{deluxetable}{lllrr}
\tablewidth{0pt}
\tabletypesize{\scriptsize}
\tablecaption{Observed lines
\label{tab-line}}
\tablehead{
\colhead{Frequency} & \colhead{}         & \colhead{}           & \colhead{$E_{l}$} \\
\colhead{(MHz)}     & \colhead{Molecule} & \colhead{Transition} & \colhead{(K)} \\
}
\startdata
465014.010   & $^{29}$SiO   & $v$=2, 11-10                & 3611    \\
428087.451  & SiO          & $v$=2, 10-9                 & 3614      \\
464244.956  & SiO          & $v$=4, 11-10                & 7086                                                   
\enddata
\tablecomments{Line information are taken from the Cologne Database for Molecular Spectroscopy (CDMS) \citep{Muller2005}. $E_{l}$ is the lower state energy.}
\end{deluxetable}

\section{Results}
\subsection{Spectral profile}
\label{sec:spec}
We successfully detected 465~GHz $^{29}$SiO $v$=2 $J$=11-10, 464~GHz SiO $v$=4 $J$=11-10, and 428~GHz SiO $v$=2 $J$=10-9 lines toward Source I. Especially, this is the first time to detect the SiO $v$=4 $J$=11-10 line in star-forming regions.
The SiO $v$=4 transition has been detected toward a red supergiant VY CMa at $J$=5-4 transition at 211~GHz for the first time~\citep{Cernicharo1993}. Recently, owing to the high spectral resolution and the sensitivity of ALMA, SiO emission with the higher vibrational excitation level has been observed in late-type stars \citep{Decin2018, Kervella2018}. 
\citet{Decin2018} detected the SiO $v$=4 and 5 $J$=8-7 line (R Dor and IK Tau) and Si$^{18}$O $v$=5 $J$=9-8 (IK Tau) in AGB stars, and \citet{Kervella2018} observed the SiO $v$=6 and 7 $J$=8-7 line in a red supergiant Betelgeuse.

\begin{figure*}[t]
\begin{center}
\includegraphics[width=5.4cm]{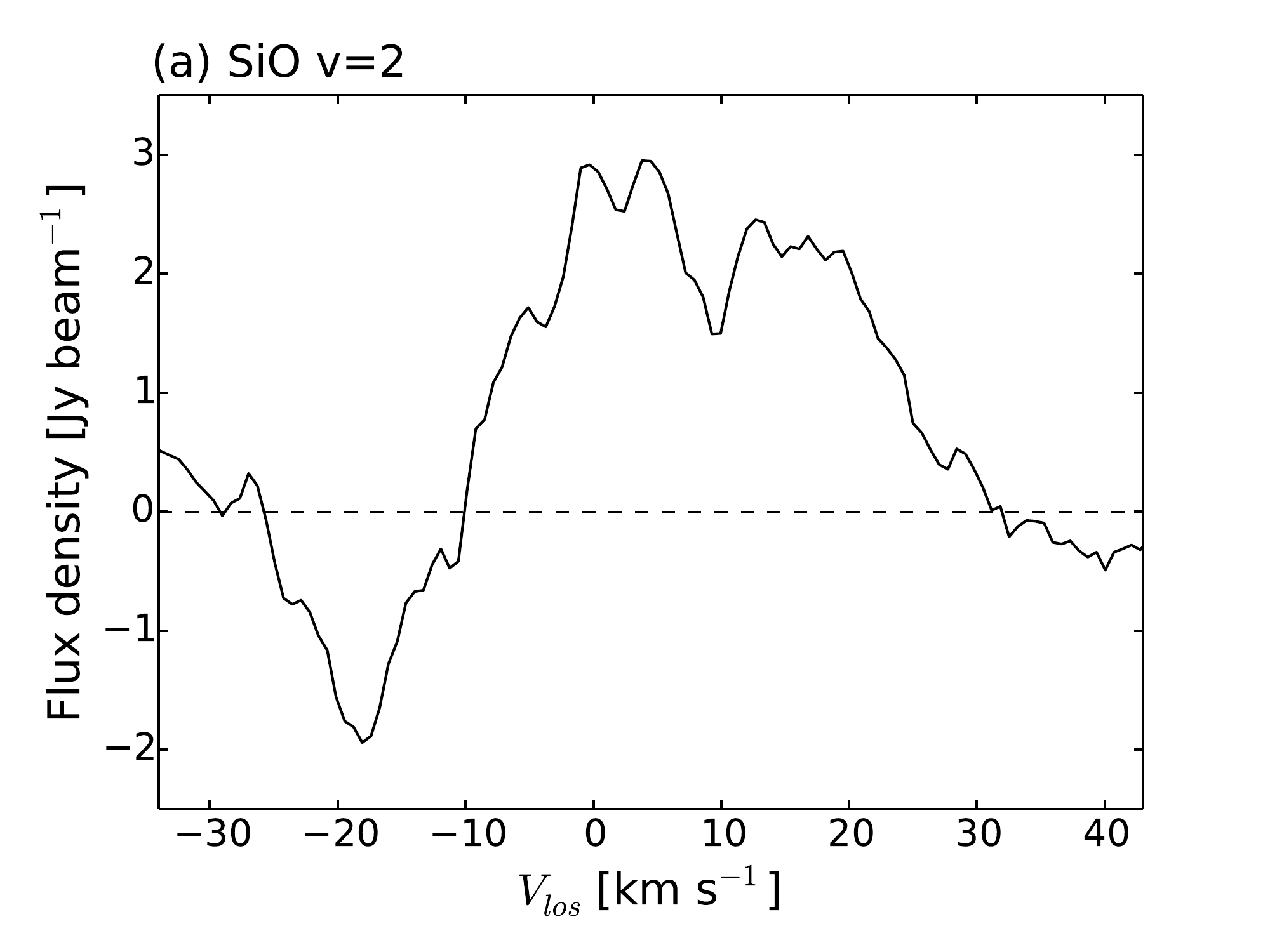}
\includegraphics[width=5.4cm]{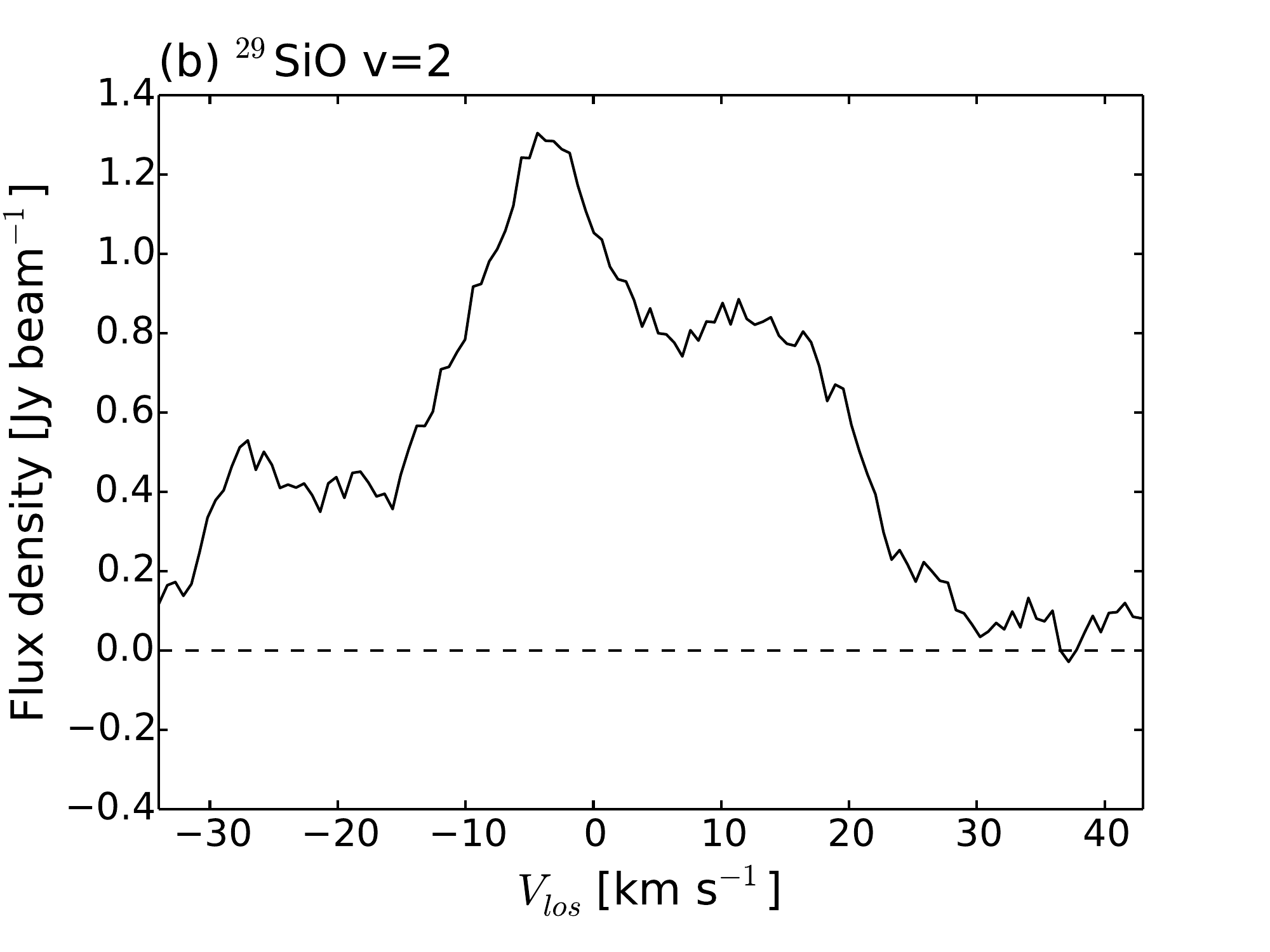}
\includegraphics[width=5.4cm]{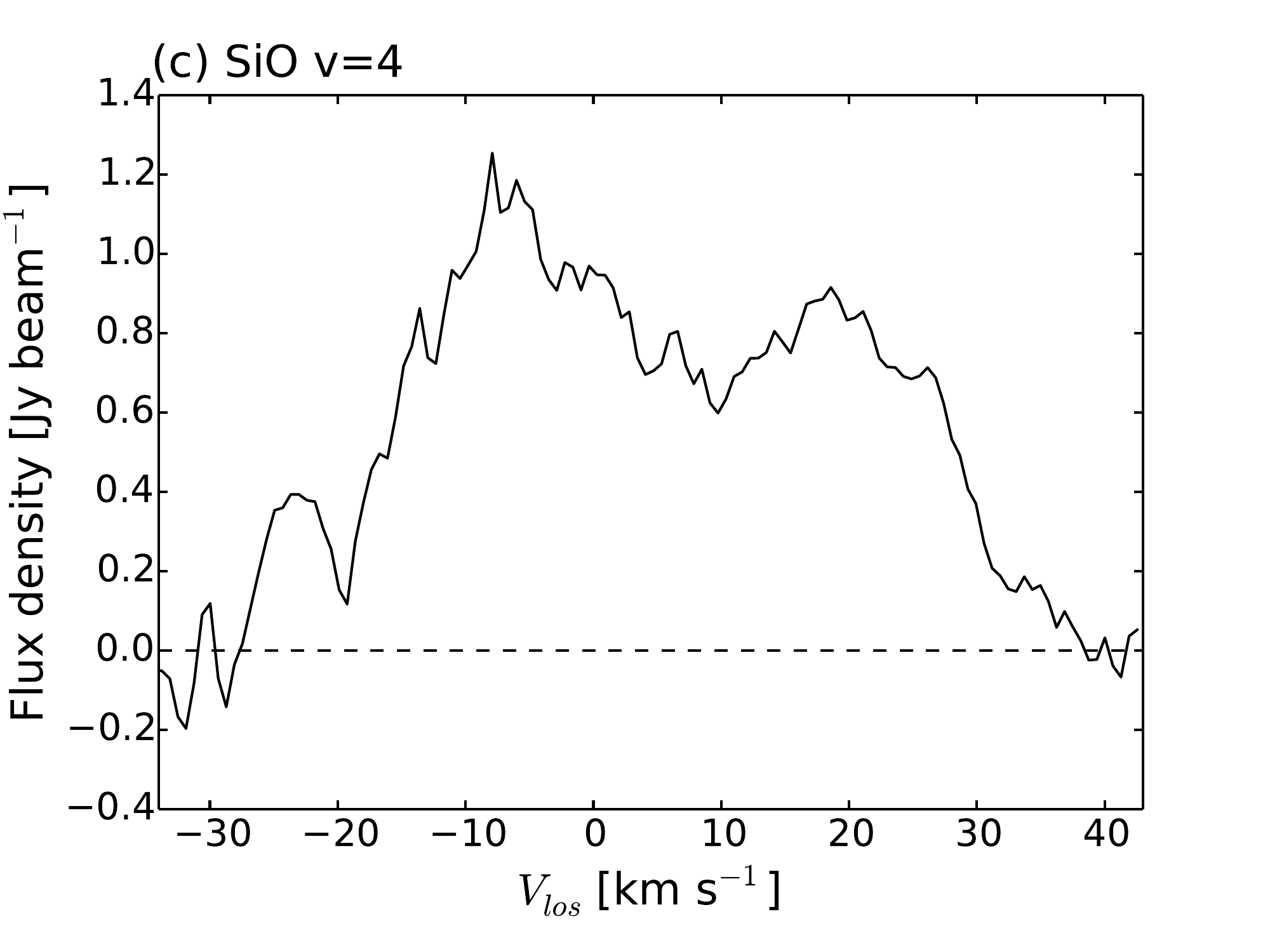}
\caption{The spectra of the observed SiO lines summed over 0.24$\times$0.24 arcsec. Horizontal dashed lines mark the baseline.
(a) The spectrum of the SiO $v$=2 $J$=10-9 line emission.
(b) The spectrum of the $^{29}$SiO $v$=2 $J$=11-10 line emission.
(c) The spectrum of the SiO $v$=4 $J$=11-10 line emission.}
\label{fig-spec}
\end{center}
\end{figure*}

Figure \ref{fig-spec} displays the observed SiO line spectra summed over 0.24\arcsec $\times$0.24\arcsec~boxes centered on the continuum peaks. The velocities of the emissions range from $\sim$-30~km~s$^{-1}$ to $\sim$30~km~s$^{-1}$, and generally centered on $V_{lsr} \sim$ 5~km~s$^{-1}$, which is the systemic velocity of Source I \citep{Plambeck1990}. The $^{29}$SiO $v$=2 $J$=11-10 and SiO $v$=4 $J$=11-10 emission show double peak profiles with the peak velocities at $\sim$-5~km~s$^{-1}$ and $\sim$15~km~s$^{-1}$, which are similar to those of other high-frequency lines and masers \citep{Kim2008, Hirota2012, Matthews2010, Greenhill2013, Hirota2014, Hirota2016a, Hirota2017, Ginsburg2018}. 

The SiO $v$=2 $J$=10-9 line spectrum also shows the double peak profile, but the velocity of the blue-shifted emission peak is shifted by $\sim$5-10~km~s$^{-1}$ from those of other emission lines. Also, the SiO $v$=2 $J$=10-9 line appears in absorption at the velocity range of -25$<V_{lsr}<$-10~km~s$^{-1}$. 
It is unlikely to be explained by a self-absorption because the SiO $v$=2 $J$=10-9 line has a high excitation temperature of $E_{l} \sim 3600$~K and they would be located very close to the central star. A possible candidate for the absorption line is the SO $v$=0 $9_{10}-8_{9}$ at 428110.747~MHz ($E_{l}=100$~K), which has a velocity of $\sim$ -1~km~s$^{-1}$ at the absorption peak. In Orion-KL, SO emission is detected in the hot core and the compact ridge gas, as well as the disk and outflow near Source I \citep{Plambeck2016, Wright1996}. However, the velocity of the gas in the hot core and the compact ridge is 2-11~km~s$^{-1}$. Thus, the absorption is likely to be associated with the SO in the outflow.
\citet{Plambeck2016} also reported the asymmetric absorption limited to blue-shifted velocities (-13$<V_{lsr}<$5~km~s$^{-1}$) in molecular lines with upper state energies $E_U<500~$K, especially in the sulfur-bearing species. They also suggested that the absorption could be associated with material within the outflow of Source I because the velocity width of the absorption corresponds to the halfwidth of other lines associated with the outflow.
 
\subsection{Spatial structure}
\label{sec:spatial}
\begin{figure*}[th]
\begin{center}
\includegraphics[width=17.5cm]{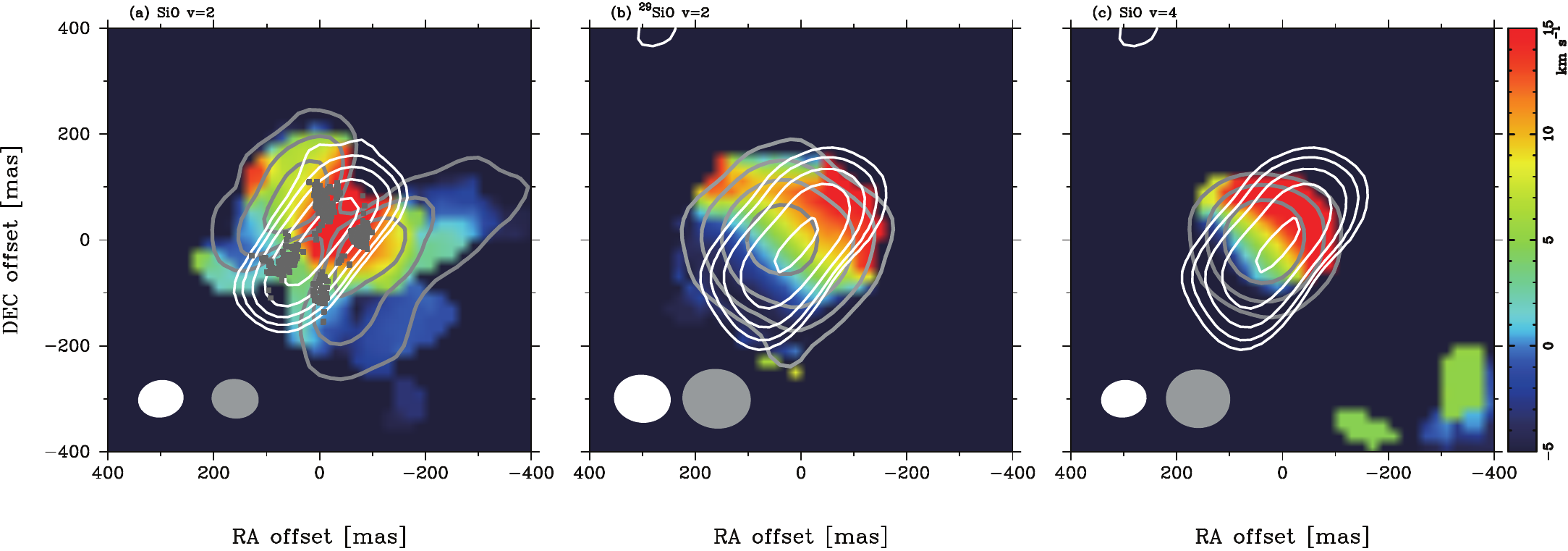}
\caption{Moment 0~(grey contour) and moment 1 maps~(color) of observed SiO lines superposed on the continuum map~(white contour). The contour levels are 5, 10, 20, 40, and 80 times root-mean-square (RMS) noise levels, and the RMS noise levels are 3~mJy~beam$^{-1}$ and 5~mJy~beam$^{-1}$, respectively, for 430~GHz and 460~GHz continuum maps. (a) Map of SiO $v$=2 $J$=10-9 and 430~GHz continuum emission. The contour levels are 5,10, and 20 times RMS noise level of 26~mJy~beam$^{-1}$ for SiO $v$=2 $J$=10-9 line. Grey dots represent the position of SiO $v$=1 and 2 $J$=1-0 maser emission \citep{Kim2008}. The beam sizes are 0.09\arcsec $\times$0.07\arcsec (grey ellipse), PA=82\degr and 0.08\arcsec$\times$ 0.07\arcsec (white ellipse), PA=-80\degr for SiO $v$=2 $J$=10-9 and 430~GHz continuum emission, respectively. (b) Map of  $^{29}$SiO $v$=2 $J$=11-10 and 460~GHz continuum emission. The contour levels are 5,10, 20, and 40 times RMS noise level of 33~mJy~beam$^{-1}$ for $^{29}$SiO $v$=2 $J$=11-10 line. The beam sizes are 0.13\arcsec$\times$0.11\arcsec, PA=82\degr (grey ellipse) and 0.10\arcsec$\times$0.09\arcsec, PA=83\degr(white ellipse) for $^{29}$SiO $v$=2 $J$=11-10 and 460~GHz continuum emission, respectively. (c) Map of SiO $v$=4 $J$=11-10 and 460~GHz continuum emission. The contour levels are 5,10, and 20 times RMS noise level of 57~mJy~beam$^{-1}$ for SiO $v$=4 $J$=11-10 line. The beam sizes are 0.12\arcsec$\times$0.11\arcsec, PA=87\degr (grey ellipse) and 0.10\arcsec$\times$0.09\arcsec, PA=83\degr (white ellipse) for SiO $v$=4 $J$=11-10 and 460~GHz continuum emission, respectively.
}
\label{fig-mom}
\end{center}
\end{figure*}

Figure \ref{fig-mom} displays the integrated intensity and peak velocity maps of SiO emissions overlaid with the continuum emission at 430~GHz and 460 GHz. 
In Figure \ref{fig-mom} (b) and (c), $^{29}$SiO $v$=2 $J$=11-10 and SiO $v$=4 $J$=11-10 emissions exhibit the compact structures centered on Source I with the deconvolved size of 0.13\arcsec$\times$0.13\arcsec~and 0.08\arcsec$\times$0.06\arcsec, respectively. Both lines show the velocity gradient along the northwest-southeast direction, which is similar with $v$=1 and 2 $J$=1-0 SiO masers \citep{Matthews2010, Kim2008} and H$_2$O emission \citep{Hirota2014, Hirota2017}.
It is clear that the $v$=4 transition with the highest excitation energy level arises at the smallest radii from the YSO, where high temperatures are expected. In contrast to the elongated structure of the continuum emission, the $^{29}$SiO $v$=2 $J$=11-10 and SiO $v$=4 $J$=11-10 emission show almost circular structure. Besides, it seems that the peak positions of the both SiO lines slightly offset to the northeast direction from the peak position of the continuum emission. 
This result is consistent with the previous ALMA observations reporting that the molecular line emission offset from the disk midplane \citep{Ginsburg2018, Plambeck2016}. \citet{Plambeck2016} reported that line emissions such as SiS are spatially extended relative to the continuum. Also, \citet{Ginsburg2018} separated the line emitting region from the continuum emission with higher angular resolution and reported that the SiO and H$_2$O lines are seen in the region below and above the continuum emission. It suggests the continuum emission is most likely optically thick emission from the disk, and molecular line emissions originate from the atmosphere of the disk and/or the base of the outflow.

The peak flux densities of $^{29}$SiO $v$=2 $J$=11-10 and SiO $v$=4 $J$=11-10 lines are 1.3~Jy and 1.2 Jy at the LSR velocity of -4.3~km~s$^{-1}$ and -6~km~s$^{-1}$, respectively.
We measured the source size of each channel as 0.2\arcsec$\times$0.16\arcsec~and 0.15\arcsec$\times$0.13\arcsec~with the two-dimensional Gaussian fitting to the channel maps. Using measured flux densities and the source sizes, the brightness temperatures of $^{29}$SiO $v$=2 $J$=11-10 and SiO $v$=4 $J$=11-10 lines are estimated to be 229~K and 349~K. The estimated brightness temperatures are lower than the temperature of the dust (500-800~K) and the gas (1000-2000~K) \citep{Hirota2016b, Plambeck2016, Goddi2009}. One of the possible explanations is that SiO lines are optically thin thermal emission.

On the other hand, we cannot exclude the possibility of the maser nature of $^{29}$SiO $v$=2 $J$=11-10 and SiO $v$=4 $J$=11-10 lines because of the similarity with other maser transitions, such as the small distribution comparable to the 43~GHz $v$=1, 2 $J$=1-0 masers, and the high density/temperature in this region. From the spectral energy distribution, \citet{Plambeck2016} inferred the disk mass of 0.02-0.2$M_{\odot}$ and the $n_H>3.2\times10^{10}$~cm$^{-3}$. The existence of SiO $v$=1 and 2 $J$=1-0 masers also suggest that the temperature and the density at the disk surface must be 1000-2000~K and $10^{10\pm1}$~cm$^{-3}$ \citep{Goddi2009}. 

In the model calculation for the excitation of SiO $v$=4 $J$=5-4 maser presented for VY CMa \citep{Cernicharo1993}, the pumping of the SiO $v$=4 maser requires the local density of $\sim$0.5-5$\times 10^{10}$~cm$^{-3}$ at the gas temperature of 1500~K. Although the physical properties of the SiO maser emitting region in a late-type star would be different from those in Source I, the disk properties of Source I is hot and dense enough to pump even $v$=4 SiO maser. In this case, the estimated brightness temperature of 229~K and 349~K would be the lower limits considering the possibility that the compact maser emission is unresolved.

In Figure \ref{fig-mom} (a), the SiO $v$=2 $J$=10-9 emission exhibits a bipolar structure centered on Source I a size of $\sim$0.5\arcsec $\times$0.5\arcsec, which is also visible in images of the low-velocity northeast-southwest outflow traced by SiO and H$_2$O lines including several maser lines \citep[e.g.,][]{Plambeck2009, Niederhofer2012, Greenhill2013, Hirota2017}. The moment 1 map clearly shows the velocity gradient in the northwest-southeast direction, and it is consistent with that of outflow with the size of $<$5\arcsec \citep{Kim2008, Greenhill2013, Hirota2017}.  
The moment 0 map indicates that the redshifted emission of the SiO $v$=2 $J$=10-9 line is dominant over the blue-shifted emission. This trend is also seen in 658~GHz and 321~GHz H$_2$O line \citep{Hirota2014, Hirota2016a}, which trace the base of the outflow from the rotation disk around Source I. We confirmed that the absorption of the blue-shifted emission is located at the southeast part of the bipolar structure. It supports the idea that the material within the outflow or the disk of Source I result in the absorption feature \citep{Plambeck2016}. 

\begin{figure*}[th]
\begin{center}
\includegraphics[width=8cm, clip=true]{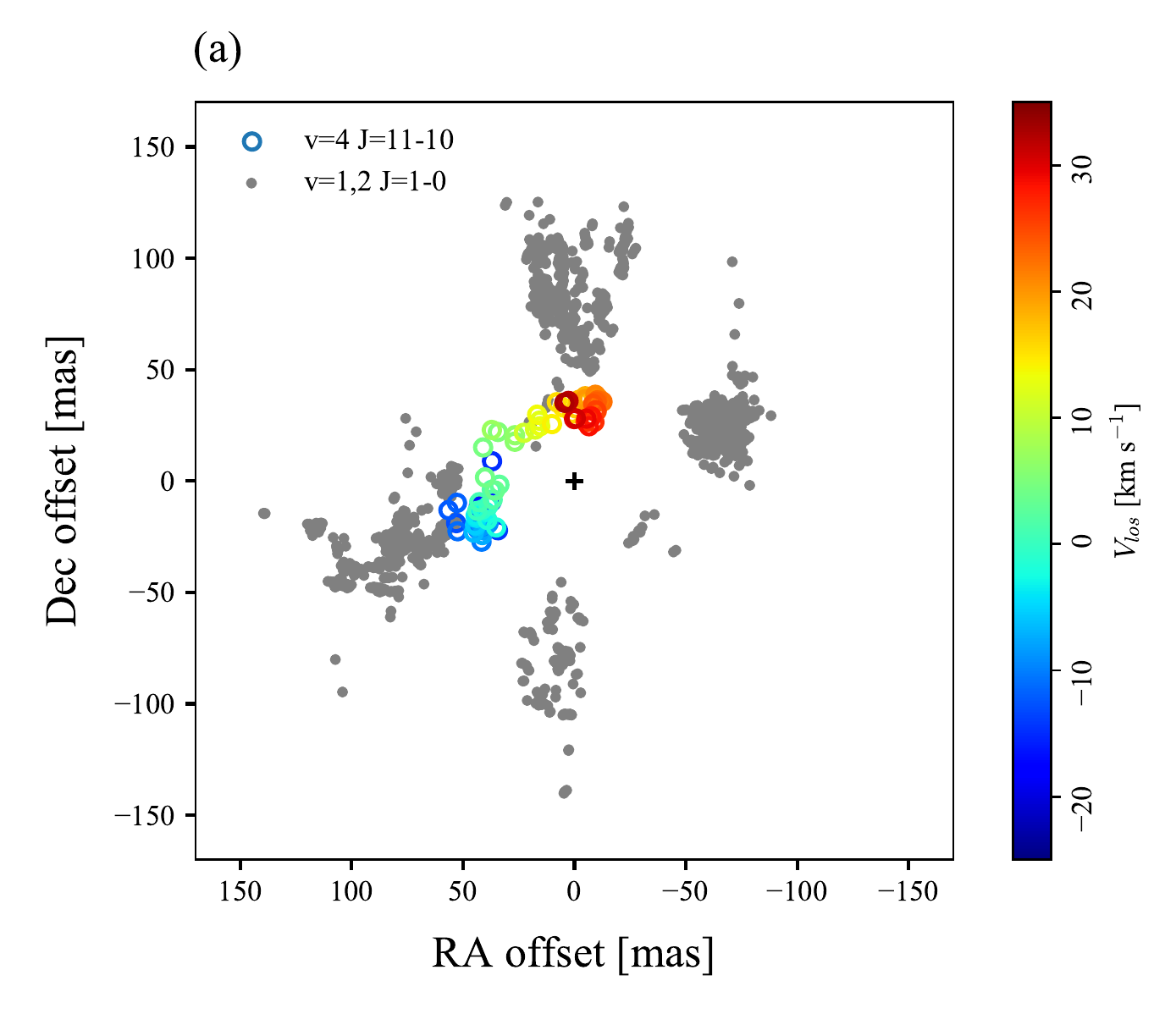}
\includegraphics[width=8cm,clip=true]{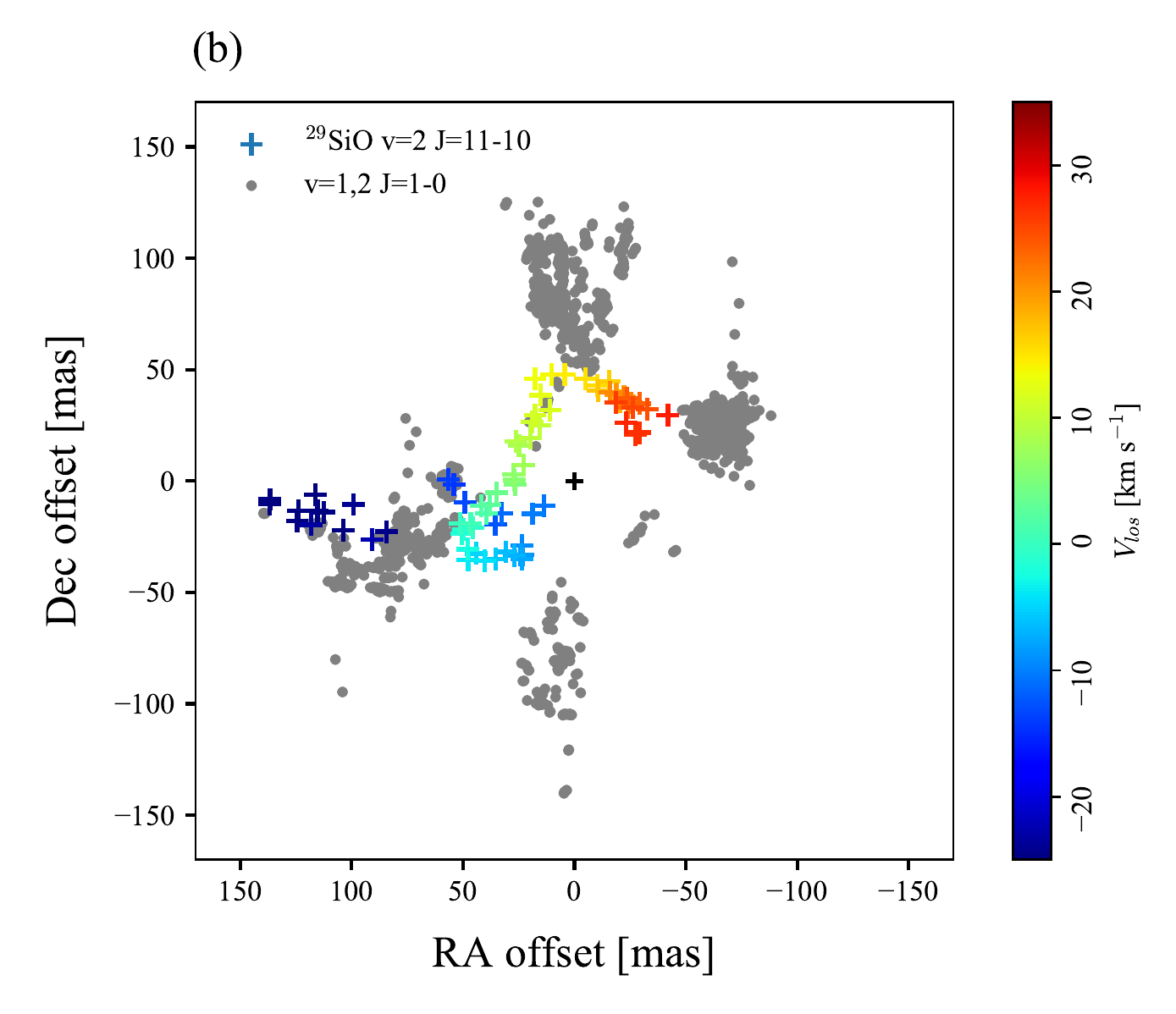}
\caption{
(a) The peak positions of the SiO $v$=1 and 2 $J$=1-0 maser emission (grey circle) and the SiO $v$=4 $J$=11-10 emission (colored circle). 
(b) The peak positions of the SiO $v$=1 and 2 $J$=1-0 maser emission (grey circle) and $^{29}$SiO $v$=2 $J$=11-10 emission (colored cross). 
The color of the SiO $v$=4 and $^{29}$SiO $v$=2 emission represents the line-of-sight velocity. The cross at (0, 0) corresponds to the position of Source I \citep{Goddi2011}.
}
\label{v124pv}
\end{center}
\end{figure*} 

\section{Discussion}
\subsection{Comparison with SiO $v$=1 and 2 $J$=1-0 masers}
Our results show that the size of the spatial distributions and the velocity ranges of high-frequency SiO emission are consistent with those of SiO $v$=1 and 2 $J$=1-0 masers. To compare the distributions between the different SiO lines, we calculated the relative positions of SiO $v$=1 and 2 $J$=1-0 masers towards the position of Source I using the absolute positions measured by \citet{Kim2008} and \citet{Goddi2011}. Then, we overlaid the maps registering the peak position of the continuum emissions to the position of Source I. Figure \ref{fig-mom} (a) shows the positions of the $v$=1 and 2 $J$=1-0 SiO masers superposed on the moment map of SiO $v$=2 $J$=10-9 line and the 430~GHz continuum emission. It reveals that the X-shaped distribution of SiO $v$=1 and 2 $J$=1-0 masers are well consistent with the base of the bipolar structure of SiO $v$=2 $J$=10-9 emission, as well as the upper and lower boundary of the continuum emission. This result confirms the idea that the SiO $v$=2 $J$=10-9 line traces the base of rotating outflow driven by the magneto-centrifugal disk wind launched by a central star \citep{Hirota2017}.

To evaluate the detailed structure of SiO $v$=4 $J$=11-10 and $^{29}$SiO $v$=2 $J$=11-10 emission, we made position centroid maps by performing two-dimensional Gaussian fitting to each velocity channel maps. 
Figure \ref{v124pv} displays the position centroids of $v$=1 and 2 $J$=1-0 SiO maser, SiO $v$=4, and $^{29}$SiO $v$=2 $J$=11-10 emissions. 
With the synthesized beam size and SNR at the peak velocity channel of each line, the positional accuracy was as high as $\sim$1~mas for the $v$=4 $J$=11-10 line and $\sim$2~mas for the $v$=2 $J$=11-10 line. The results demonstrate the offset of $v$=4 $J$=11-10 line and $^{29}$SiO $v$=2 $J$=11-10 emissions from the disk midplane, as also seen in Section~\ref{sec:spatial}. While \citet{Ginsburg2018} resolved the molecular line emission from above and below the disk, we could not detect the emission in the southwest of the disk, possibly due to the lower sensitivity and resolution. 

The $^{29}$SiO $v$=2 $J$=11-10 emission is mainly found between the eastern and northern arms of X-shape, and also in the vertical structures from the disk, such as blue-shifted emission along the eastern arm and high-velocity components near the disk midplane. It confirms that the $^{29}$SiO $v$=2 $J$=11-10 emission traces the rotating outflow as other SiO line emission \citep[e.g.]{Plambeck2016, Hirota2017, Ginsburg2018}. On the other hand, the $v$=4 $J$=11-10 emission only found in a line connecting the base of eastern and northern arms of 43~GHz SiO masers, which resembles the H$_2$O emission in \citet{Ginsburg2018}. The morphology and velocity gradient of this line suggests that it traces both the surface of the disk and the base of the outflow. 

\subsection{The kinematics of $v$=4 $J$=11-10 line emission }
\begin{figure*}[t]
\begin{center}
\includegraphics[width=7.9cm, clip=true]{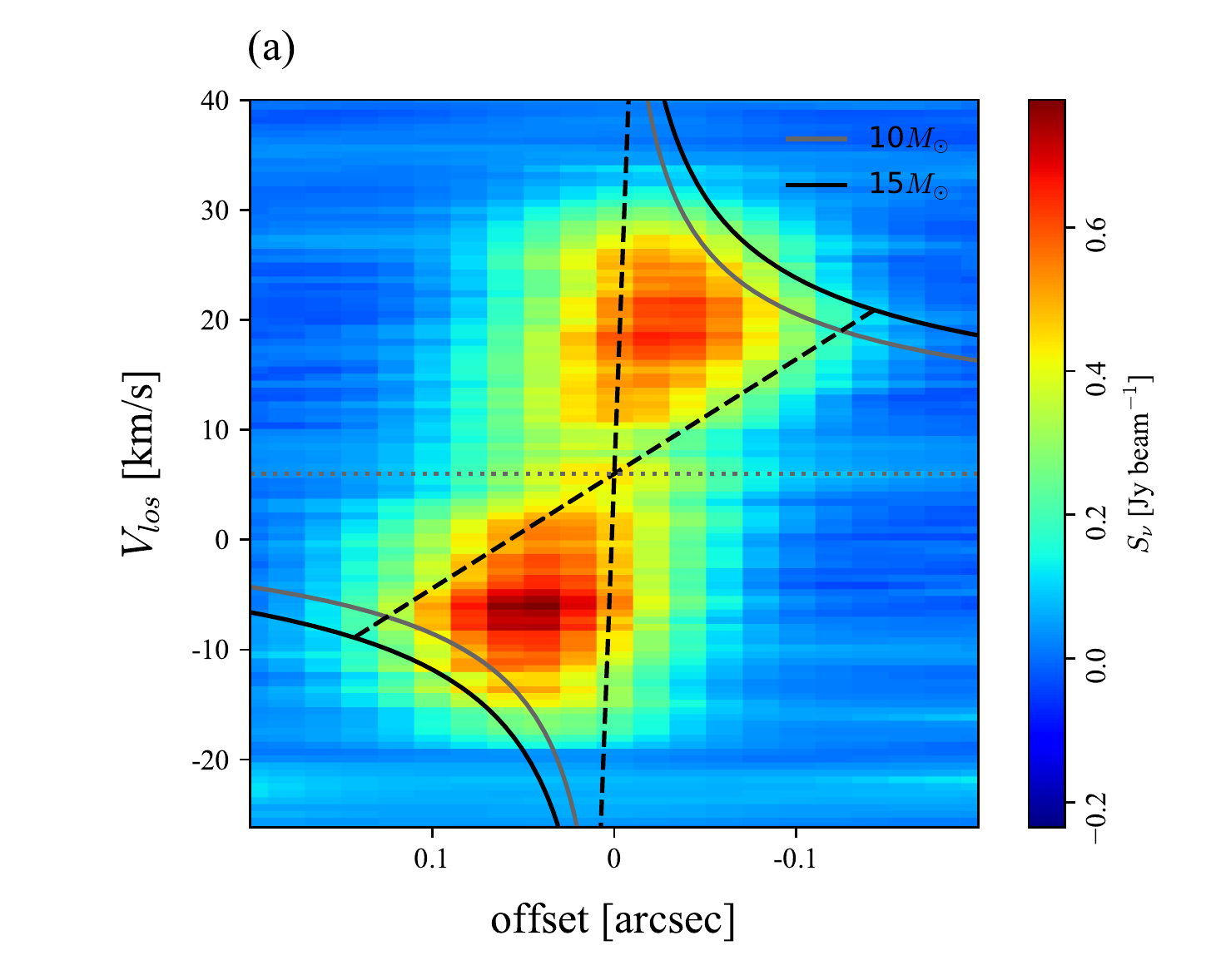}
\includegraphics[width=8.4cm,clip=true]{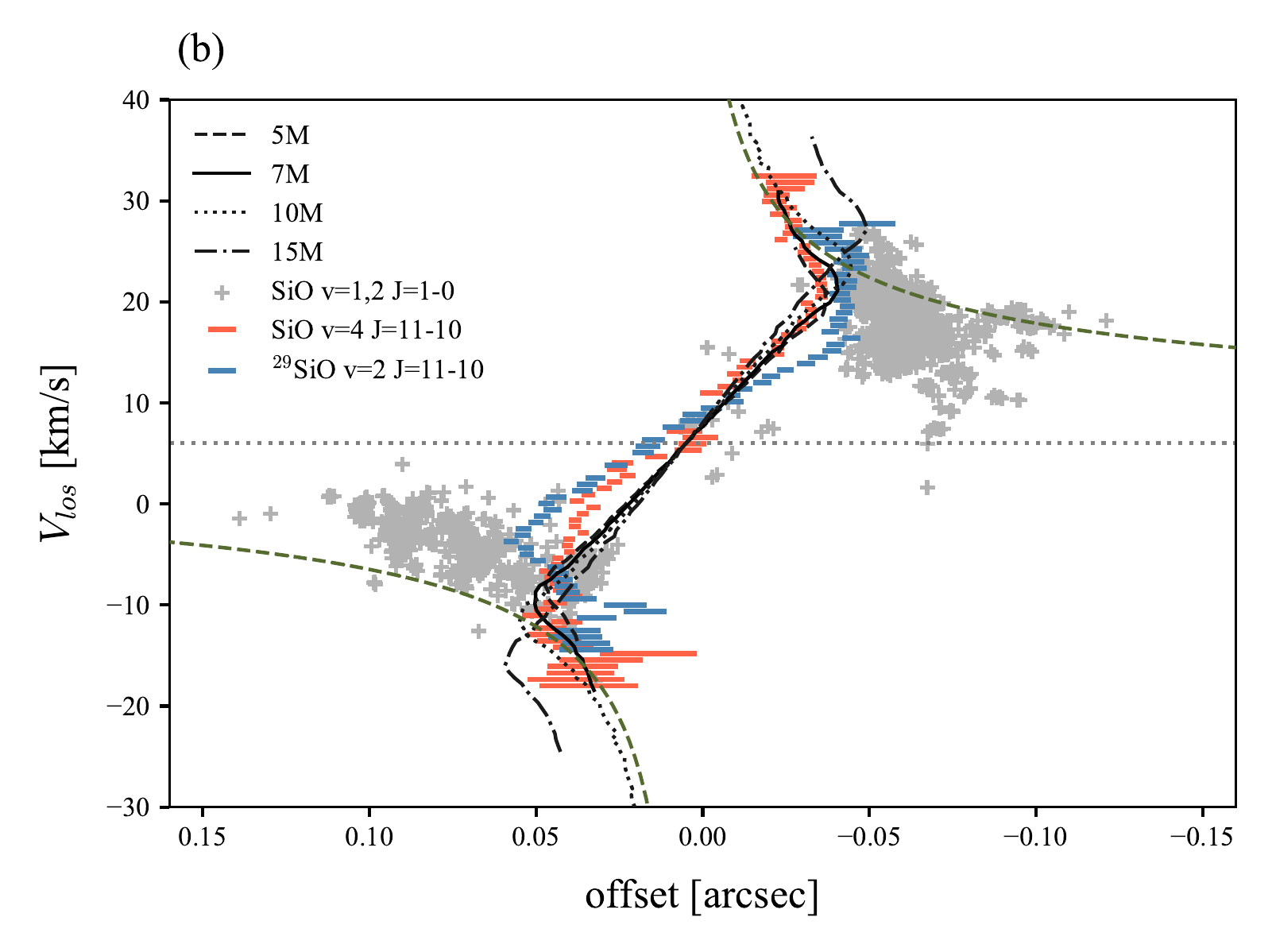}
\caption{
(a) The position-velocity diagram of the SiO $v$=4 $J$=11-10 emission. The Keplerian velocity curves for 10$M_{\odot}$ (grey) and 15$M_{\odot}$ (black) are overlaid. The dashed lines show the velocity curves at 5~AU and 60~AU. The dotted horizontal line represents the systemic velocity of 6~km~s$^{-1}$.
(b) The position-velocity diagram using the centroids of each velocity channel of the SiO $v$=1 and 2 $J$=1-0 SiO maser emission (grey cross), the $v$=4 $J$=11-10 (orange) and $^{29}$SiO $v$=2 $J$=11-10 (blue) emission. The solid line indicates the position-velocity diagram of the ring model with parameters of $M = 7M_{\odot}$, $r_{in}$=12~AU, $r_{out}$=26~AU. Also, we overlaid the model curves with $M = 5M_{\odot}$ with $r_{in}$=13~AU, $r_{out}$=23~AU, $M = 10M_{\odot}$ with $r_{in}$=6~AU, $r_{out}$=30~AU, and $M=15M_{\odot}$ with $r_{in}$=16~AU, $r_{out}$=30~AU. The dashed green line represents the Keplerian rotation curve with $M = 7M_{\odot}$. The horizontal dashed line marks the adopted systemic velocity of 6~km~s$^{-1}$.
}
\label{pv}
\end{center}
\end{figure*}

Figure \ref{pv} shows the position-velocity diagrams of the SiO $v$=4 $J$=11-10 and $^{29}$SiO $v$=2 $J$=11-10 emission along the axis with the position angle of 140\degr \citep{Hirota2014}. The spatial distribution and the velocity gradients along the major axis of the continuum emission indicate that SiO $v$=4 $J$=11-10 emission is related to the rotating disk. However, the absence of the high-velocity components at the central position suggests that the high-frequency SiO line emissions are associated with a rotating ring structure rather than a filled disk. To find the parameters of the ring structure, we fit the observed rotation curve with a simulated rotation curve using a simple model similar to those of \citet{Hirota2014} and \citet{Plambeck2016}.
We assumed a thin, edge-on ring structure in Keplerian rotation around a central star. In each position, the gas with uniform temperature and density emit the optically thin emission. The thermal width of 3~km~s$^{-1}$ was assumed. We set the systemic velocity of 6~km~s$^{-1}$, which is the central value of the observed velocity range. We simulated the rotation curve for a range of inner radii $r_{in}$ and outer radii $r_{out}$ with the central masses ranges from 5 to 15$M_{\odot}$. As a result, we assess that $r_{in}$=12~AU, $r_{out}$=26~AU, and the central mass of 7$M_{\odot}$ for the best parameter. 

Figure \ref{pv}~(b) presents the results of the fitting to the velocity centroids of the SiO $v$=4 $J$=11-10 emission. We also display the rotation curves with the central mass of 5, 10, and 15$M_{\odot}$. The rotation curves for smaller masses also seems to fit with the observed rotation curve near the systematic velocity, but it could not reproduce the high-velocity components at the innermost radii. In contrast, a simulation with a higher mass requires the smaller inner radii and larger outer radii to fit the observed rotation curve. However, we cannot exclude the possibility of the higher central mass because we did not consider the offset from the disk midplane and the effect of the outflow. Also, it is possible that the spatial resolution and sensitivity of our observation was not enough to detect the detailed structure near the inner and outer radii. Thus, we notice again that the mass of 7$M_{\odot}$ is a lower limit of the mass of Source I. 

The estimated mass of 7$M_{\odot}$ is consistent with those derived by \citet{Plambeck2016} and \citet{Hirota2014}, as well as the mass estimated with 43~GHz SiO masers \citep{Matthews2010}. The velocity gradient of the $v$=1 and 2 SiO masers supports the rotating structure with the enclosed mass of $\sim$7-8$M_{\odot}$ \citep{Kim2008, Matthews2010}. The rotation curve of 336~GHz H$_2$O and 340~GHz molecular lines also suggest the rotating ring with the central mass of $\sim$5-7$M_{\odot}$ \citep{Hirota2014, Plambeck2016}. Also, \citet{Hirota2017} suggested the mass of Source I to be 8.7$M_{\odot}$ with the rotation of the Si$^{18}$O $J$=12-11 emission.

On the other hand, recent ALMA observation with the resolution $<0.06$\arcsec detected 232~GHz H$_2$O line tracing the X-shape distribution of SiO $v$=1 and 2 $J$=1-0 masers \citep{Ginsburg2018}. Using the outer edge of the position-velocity diagram, they estimated a mass of Source I of $>$ 15$\pm$2~$M_{\odot}$, which is twice of the previous estimations and the result of this observation. 
This inconsistency could be caused by the fact that we missed a faint emission at high-velocity component near the central source due to the insufficient resolution and sensitivity. 
Also, because our model based on the emission centroids samples only the peak positions at the position-velocity diagram, it gives the lower mass compared with the case using the highest velocity at a certain position as \citet{Ginsburg2018} does.

Figure \ref{pv}~(a) displays the position-velocity diagram of the SiO $v$=4 $J$=11-10 emission. We could not fit the data with the Keplerian rotation curve because the power-law profile is not clear in our PV diagram, so we overlaid the Keplerian velocity curve for 10 and 15$M_{\odot}$. It also shows the possibility that the estimated mass can be modified if we can confine the velocity structure more precisely. The future observations with the higher resolution and sensitivity are required to constrain the central mass with the kinematics in this scale.  

Although the resolution of our observation is not enough to resolve the issue on the mass estimation of Source I, our observation shows that the SiO $v$=4 $J$=11-10 emission traces the small region with radius $<$30~AU tracing the disk surface and the inner outflow. Their high excitation level and the compact distribution implies that SiO $v$=4 $J$=11-10 emission is a promising tracer to study the kinematics of the innermost region where the outflow is launching with the higher resolution observations in the future.

\subsection{Summary}
We observed SiO line around Source I with a resolution of $\sim$0.1\arcsec and found the following:\\
1. We detected SiO $v$=4 $J$=11-10 line near Source I. This is the first detection of SiO $v$=4 emission in the star-forming regions. Besides, the nature of the emission enables us to trace the inner part of the disk/outflow system of Source I using observations with higher resolution in the future.  \\
2. SiO $v$=2 $J$=10-9 emission shows the bipolar distribution with a size of 0.5\arcsec$\times$0.5\arcsec, tracing the base of the northeast-southwest outflow from Source I. While, $^{29}$SiO $v$=2 and SiO $v$=4 $J$=11-10 line show the compact distribution tracing the surface of the disk and the base of the outflow of Source I.\\
3. The centroids of the velocity distribution of SiO $v$=4 $J$=11-10 line suggest that the emission traces the ring structure with 12~AU $<r<$ 26~AU and the central mass of 7$M_{\odot}$. The estimated value would be a lower limit of the mass because the sensitivity and resolution of our observation are not enough to resolve the faint components in the rotation curve. \\

\acknowledgements
This paper makes use of the following ALMA data: ADS/JAO.ALMA\#2013.1.00048.S. 
ALMA is a partnership of ESO (representing its member states), NSF (USA) and NINS (Japan), together with NRC (Canada), NSC and ASIAA (Taiwan), and KASI (Republic of Korea), in cooperation with the Republic of Chile. The Joint ALMA Observatory is operated by ESO, AUI/NRAO and NAOJ. 
We thank the staff at ALMA for making observations and reducing the data. 
TH is financially supported by the MEXT/JSPS KAKENHI Grant Numbers 24684011, 16K05293, 17K05398, and 18H05222. 
KM is supported by the MEXT/JSPS KAKENHI Grant Number 15K17613. 
MH is supported by the MEXT/JSPS KAKENHI Grant Numbers 24540242 and 25120007. 
Data analysis were in part carried out on common use data analysis computer system at the Astronomy Data Center, ADC, of NAOJ. 

{\it Facilities:} \facility{ALMA}.

{}
\end{document}